# UV Surface Habitability of the TRAPPIST-1 System

Jack T. O'Malley-James[1*] and L. Kaltenegger[1]

*Carl Sagan Institute, Cornell University, Ithaca, NY 14853, USA*
**jto28@cornell.edu*

**ABSTRACT**

With the discovery of rocky planets in the temperate habitable zone (HZ) of the close-by cool star TRAPPIST-1 the question of whether such planets could harbour life arises. Habitable planets around red dwarf stars can orbit in radiation environments that can be life-sterilizing. UV flares from these stars are more frequent and intense than solar flares. Additionally, their temperate HZs are closer to the star. Here we present UV surface environment models for TRAPPIST-1's HZ planets and explore the implications for life. TRAPPIST-1 has high X-ray/EUV activity, placing planetary atmospheres at risk from erosion. If a dense Earth-like atmosphere with a protective ozone layer exists on planets in the HZ of TRAPPIST-1, UV surface environments would be similar to present-day Earth. However an eroded or an anoxic atmosphere, would allow more UV to reach the surface, making surface environments hostile even to highly UV-tolerant terrestrial extremophiles. If future observations detect ozone in the atmospheres of any of the planets in the HZ of TRAPPIST-1, these would be interesting targets for the search for surface life. We anticipate our assay to be a starting point for in-depth exploration of stellar and atmospheric observations of the TRAPPIST-1 planets to constrain their UV-surface-habitability.

## 1 INTRODUCTION

The ultra-cool dwarf star TRAPPIST-1 hosts seven Earth-sized planets (Gillon et al. 2017, Gillon et al. 2016). Two are closer to the star than the inner-edge of the star's HZ, one is between the empirical and 3D inner HZ boundaries (Kopparapu et al. 2014, Leconte et al. 2013) and one is outside the outer-edge of the empirical HZ. The remaining three, TRAPPIST-1e, -f and -g, have orbits that place them within the HZ (Fig. 1). More habitable planet discoveries in the near future are likely to be found orbiting nearby cool stars – for example, K2 could detect TRAPPIST-like planets for about 10% of M stars sampled (Demory et al. 2016) and the upcoming NASA TESS mission will detect dozens of rocky planets in the HZs of cool stars (Sullivan et al. 2015). Therefore, it is important to characterise how the surface UV environments of these planets could affect biology.

Recent work is beginning to unravel some of the factors affecting the habitability of the TRAPPIST-1 system (Bolmont et al. 2017). X-ray observations show the star is a strong, variable coronal X-ray source, with a similar X-ray luminosity to the quiet Sun (Wheatley et al. 2017). As TRAPPIST-1 is only about 5% as luminous as the Sun, HZ planets are much closer to their host star than Earth is to the Sun and thus would be subject to significant X-ray/EUV fluxes, which could alter their primary and secondary atmospheres (Wheatley et al. 2017). Recent Lyman-α observations of the system suggest that TRAPPIST-1's chromosphere is moderately active, which could cause the inner planets to lose their atmospheres within a few billion years (Bourrier et al. 2017). However the UV-activity (flare frequency and intensity) of TRAPPIST-1 in the biologically-relevant 100-400 nm range is unknown.

Currently, the only information we have on the atmospheres of the TRAPPIST-1 planets comes from a combined transmission spectrum for the two inner planets (TRAPPIST-1c and d) that shows a lack of features, ruling out cloud-free hydrogen-dominated atmospheres for planets b and c, but leaving a wide range of possible atmospheres open (De Wit et al. 2016). This allows for a wide range of potential UV surface radiation environments.

As UV radiation can lead to sterile surface conditions (France et al. 2013, Wheatley et al. 2017, Buccino, Lemarchand & Mauas 2007, Segura et al. 2010), we therefore explore how the surface UV radiation environment would change with different planetary atmospheres and for high and low stellar UV activity levels. On Earth, only UV-A (315-400 nm) reaches the surface due to the protection of the ozone layer; however, different types of atmospheres and stellar activity levels can cause more biologically harmful higher energy UV-B (280-315 nm) and UV-C (100-280 nm) radiation to reach planetary surfaces.

We base our model planets on the three HZ planets, which receive stellar fluxes of approximately 0.66, 0.38 and 0.26 times the flux on Earth ($S_0$). A star's radiation shifts to longer wavelengths with cooler surface temperatures, which makes the light of a cooler star more efficient at heating an Earth-like planet (Kasting, Whitmire & Reynolds 1993). This is partly due to the effectiveness of Rayleigh scattering, which decreases at longer wavelengths. A second effect is the increase in near-IR absorption by $H_2O$ and $CO_2$ as the star's spectral peak shifts to these wavelengths. That means that the integrated stellar flux that hits the top of a planet's atmosphere from a cool red star warms a planet more efficiently than the same integrated flux from a hot blue star. Therefore a planet around TRAPPIST-1 requires less stellar irradiation than Earth for comparable surface temperatures. The effectiveness of individual stars at heating an Earth-like planet's surface (given as weighting factors) can be found in Kaltenegger & Haghighipour (2013).

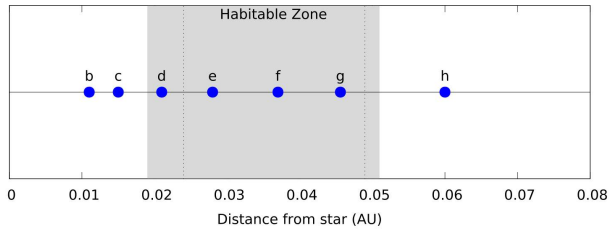

**Figure 1.** *The habitable zone of the TRAPPIST-1 system. The most likely positions of the known planets are plotted in blue. Conservative HZ limits are shown by dashed lines. The fluxes for the HZ planet positions are used for the model planets in our study. These receive fluxes of 0.258, 0.382 and 0.662 times the solar flux at Earth, respectively (Gillon et al. 2017).*

Current stellar models are unable to model the UV region from M dwarfs self-consistently (France et al. 2013, Rugheimer et al. 2015a) making observations of cool stars in the UV a critical input for habitability calculations. In lieu of UV data for TRAPPIST-1, we use two stellar input spectra (Rugheimer et al. 2015a), which represent two possible extremes of UV activity for a star like TRAPPIST-1 – (i) an "active" spectrum based on the most active M dwarf UV measurements and (ii) an "inactive" spectrum based on a semi-empirical model without a chromosphere, constructed to represent the lowest theoretical UV flux (see Fig. 2) and compare their results.

## 2 METHODS

We use a coupled climate-photochemistry code to determine the UV surface flux (see Segura et al. 2007, Rugheimer et al. 2015b for details) for our model planets within TRAPPIST-1's HZ. The high X-ray/EUV flux from TRAPPIST-1 (and other similarly active young M stars) should strip away a planet's atmosphere over time (Gillon et al. 2016), which would lead to less dense atmospheres and/or lower $O_2/O_3$ levels than on Earth. Therefore we model UV planetary surface fluxes under three atmospheric scenarios: (i) a present-day Earth-like oxygen atmosphere, (ii) a thin oxygen atmosphere (0.1 bar), (iii) a 1 bar $CO_2$-dominated atmosphere similar to the early Earth 3.9 Gyr ago (following Kaltenegger, Traub & Jucks 2007, Rugheimer et al 2015).

To model these planetary atmospheres, we use a coupled 1D radiative-convective atmosphere code developed for rocky exoplanets, EXO-Prime (see Kaltenegger & Sasselov 2010 for details), based on an iterative photochemistry code (Kasting & Ackerman 1986, Kasting, Holland & Pinto 1985, Toon et al. 1989, Segura et al. 2007, Kopparapu et al. 2014) that solves for 55 chemical species linked by 220 reactions using a reverse-Euler method. A two-stream approximation (Toon et al. 1989), including multiple scattering by atmospheric gases, is used in the visible/near IR to calculate shortwave fluxes. Four-term, correlated-k coefficients parametrise the absorption. Exo-Prime (a detailed description of the code can be found in Kaltenegger et al. 2010) calculates the atmospheric temperature as well as chemical make-up depending on stellar irradiation and outgassing rates (see Fig.3). The average 1D global atmospheric model profile is calculated down to a set pressure that corresponds to a certain height using a plane-parallel atmosphere, treating the planet as a Lambertian sphere, and setting the stellar zenith angle to 60° to represent the average incoming stellar flux on the dayside of the planet (see also Schindler & Kasting 2000). We use present-day Earth outgassing rates for atmospheres with oxygen: The biogenic (produced by living organisms) fluxes were held fixed in the models in accordance with the fluxes that reproduce the modern mixing ratios in the Earth–Sun case, except for $CH_4$ and $N_2O$, which were given a fixed mixing ratio of $1.0 \times 10^{-3}$ and $1.5 \times 10^{-2}$, respectively (following Segura et al. 2005). The anoxic atmosphere is described in detail in Kaltenegger et al. (2007). We model clear sky atmospheres for all planets. We do not include clouds or hazes in our model because the effect of clouds or hazes on surface UV radiation is unclear. Clouds can both block or focus incoming UV radiation to the ground depending on cloud cover and type (Grant & Heisler 1997; Parisi & Downs 2004).

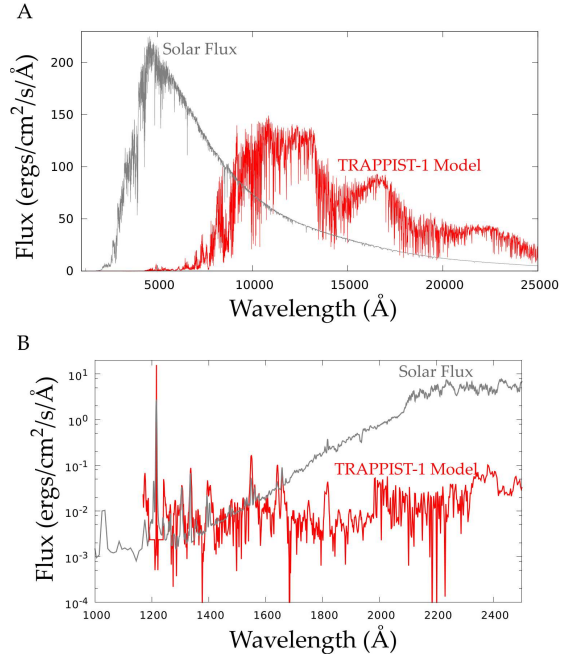

**Figure 2. A.** *The model stellar spectrum used in our simulations. The solar spectrum is plotted in grey for comparison.* **B.** *The near UV portion of the spectra. The UV-active model spectrum for TRAPPIST-1 overlaps with, or exceeds the high energy solar UV-C spectrum. Both spectra show the flux at a 1-AU-equivalent distance.*

## 3 RESULTS & DISCUSSION

The atmospheric profiles and chemical mixing ratios for three key gases, $H_2O$, $O_3$ and $CH_4$ in the planetary atmosphere models are shown in Fig.3 and UV irradiation and UV surface environments in Fig.4, colour coded by HZ planet (e = green, f = magenta, g = blue).

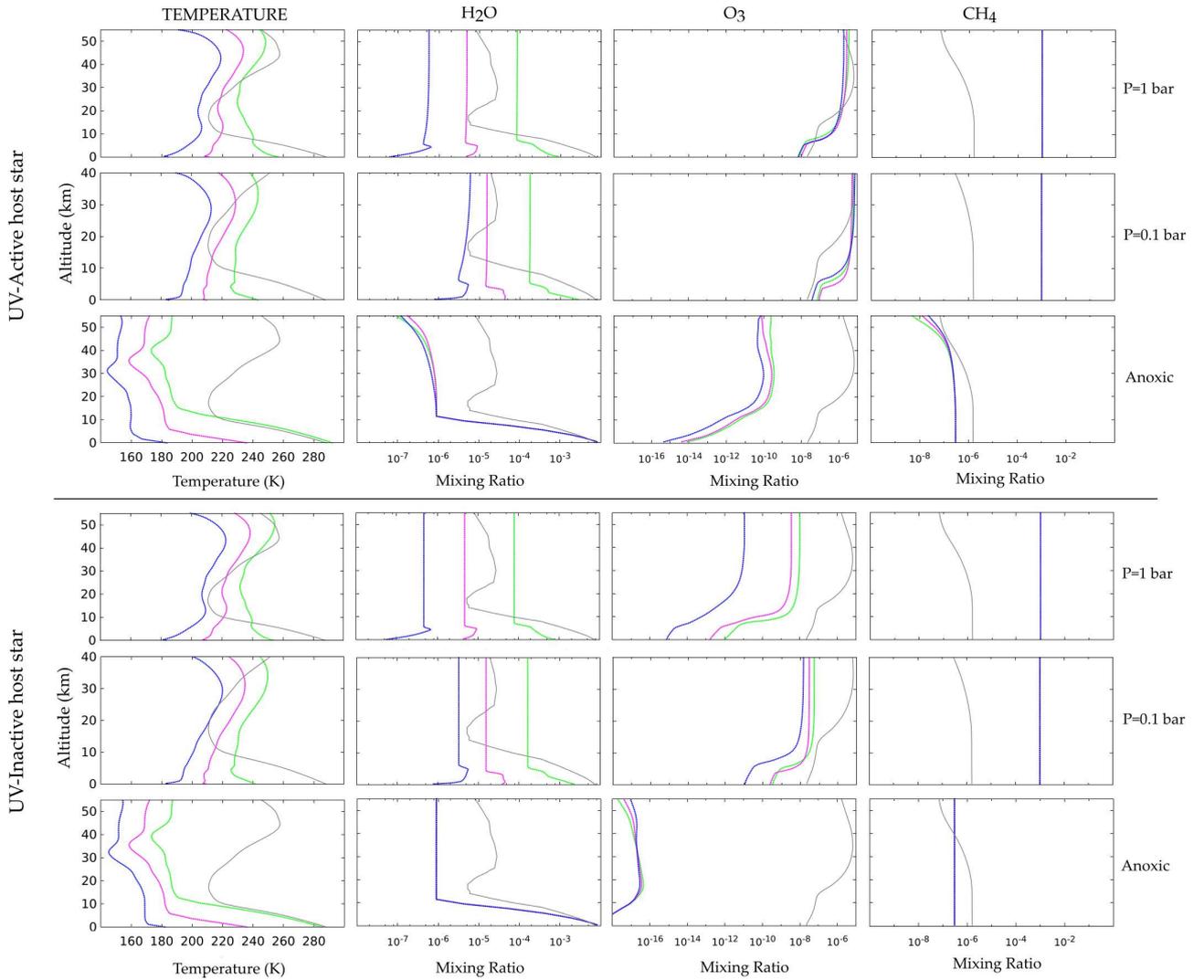

**Figure 3.** *The temperature-pressure profiles, ozone, water and methane mixing ratios for each of the atmosphere cases modelled for the three planets in the HZ of TRAPPIST-1 (-e (green), -f (magenta), -g(blue)). Note the atmospheric height is lower for the eroded atmosphere.*

The surface temperature of the planet models decreases with increasing orbital distance, due to less total stellar irradiation. For eroded atmosphere models, the surface temperatures of the planets further decrease compared to the 1 bar case due to decreased greenhouse effect. However, for the outer two HZ planets this decrease is small as a result of greenhouse heating effects ceasing to be effective at such low stellar irradiation for low pressure atmospheres.

For the oxygen-rich model atmospheres, $H_2O$ mixing ratios decrease with increasing distance from the star. Notable inversions form near the surface for the colder planets (f and g) as a result of low stellar flux and surface heat-loss processes. The eroded atmosphere planet models show higher $H_2O$ mixing ratios compared to the 1 bar atmospheres as a result of the decreased boiling point of water at lower pressures. UV fluxes shortward of 240 nm, which are higher in the active star model, drive the photodissociation of oxygen, driving $O_3$ production. Ozone mixing ratios for high stellar UV activity are similar to Earth's. For eroded atmospheres, the maximum $O_3$ concentration occurs at lower altitudes due to decreased atmospheric pressure. The $O_3$ mixing ratio is similar in the 1 bar and eroded atmospheres, but the overall $O_3$ column depth is lower in eroded atmospheres due to the lower total atmospheric mass. $CH_4$ mixing ratios are higher in these models than on Earth because M stars emit lower fluxes in the 200-300 nm range that drives $CH_4$ photodissociation (Segura et al. 2005), giving $CH_4$ a longer atmospheric lifetime than on Earth.

For an anoxic atmosphere, no significant ozone layer develops for either stellar model. The cooler upper atmosphere temperature profiles in the anoxic atmospheres are a result of a combination of lower high-energy UV fluxes from the star (compared to Earth) and low atmospheric $CH_4$ levels, reducing atmospheric heating (see e.g. Segura et al. 2005; Rugheimer et al. 2015, Meadows et al. 2016). High stellar UV activity decreases upper atmosphere $H_2O$ and $CH_4$ mixing ratios, more pronounced the closer the planet is to the star. The UV flux from inactive star models is too low for any notable change in the $H_2O$ or $CH_4$ mixing ratio with orbital distance.

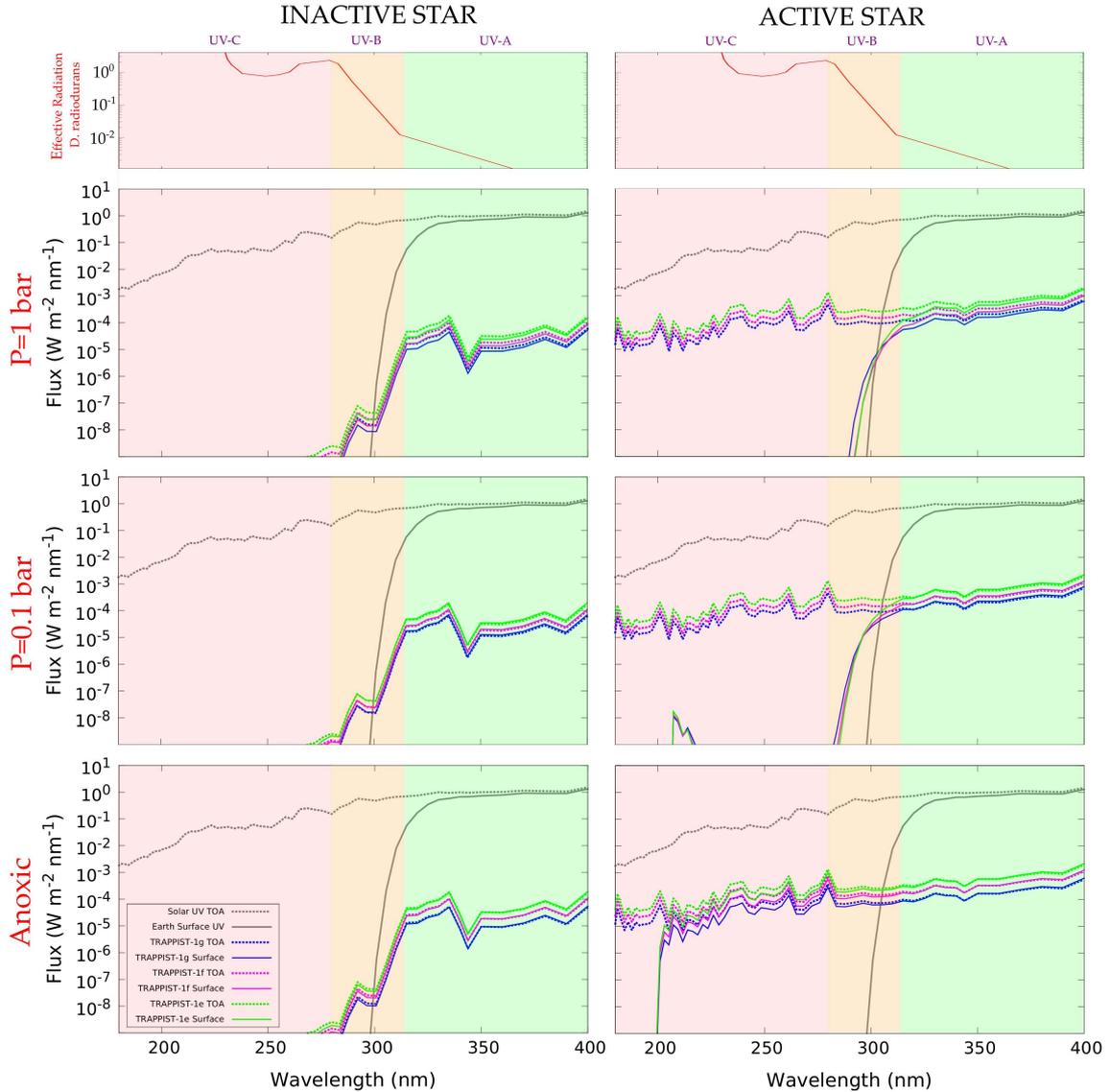

**Figure 4.** *The UV top-of-atmosphere and surface fluxes for the model planets in the TRAPPIST-1 HZ. The top panels show the UV action spectrum (normalised to 1 at ~260 nm) for the highly radiation-tolerant microorganism D. radiodurans. Planets with stellar incident fluxes of 0.66 $S_0$ , 0.38 $S_0$ and 0.26 $S_0$ are modelled (1 $S_0$ represents the equivalent flux received by Earth from the Sun). The UV flux (surface and top-of-atmosphere) for the present day Earth is plotted for comparison (grey lines). The left-hand column shows the results for a UV-inactive model of TRAPPIST-1. The right-hand column shows the results for a UV-active model of TRAPPIST-1 for planets with Earth-like atmospheres, low density (0.1 bar) Earth-composition atmospheres, and high-$CO_2$ early Earth-like atmospheres.*

A planet's atmospheric composition influences the surface UV environment. Fig.4 shows the stellar UV irradiance at the top of the planet's atmosphere (TOA, dotted lines) and the resulting surface UV (solid lines) of the three planets for different atmospheric compositions as well as the present-day Earth UV surface environment (grey solid line). To assess the damage a UV surface environment causes to life as we know it, we also show the action spectrum describing the relative mortality rates at different UV wavelengths of the radiation-tolerant extremophile *Deinococcus radiodurans* (Calkins & Barcelo 1982, Setlow & Boling 1965) (red line) and indicate the different UV regimes (A, B and C), with increasing biological harmfulness respectively. *Deinococcus radiodurans* is one of the most radiation resistant organisms known on Earth (Rothschild & Mancinelli 2001). Therefore, we use this as a benchmark against which to compare the habitability of the different radiation models. The action spectrum compares the effectiveness of different wavelengths of UV radiation at inducing a 90% mortality rate. It highlights which wavelengths have the most damaging irradiation for biological molecules: For example, the action spectrum in Fig. 4 shows that a dosage of UV radiation at 360 nm would need to be three orders of magnitude higher than a dosage of radiation at 260 nm to produce similar mortality rates in a population of this organism. Note that the action spectrum is only plotted to 230nm. However, the biological effectiveness of UV-C radiation < 230 nm should track the UV absorption profile of DNA (in the absence of cellular radiation screening) and so will continue to have increasingly lethal effects on exposed organisms (see e.g. Lindberg & Horneck 1991).

Shortwards of 200 nm atmospheric $CO_2$ and $H_2O$ absorb UV radiation, blocking it from reaching the ground in all our models. Ozone absorbs UV radiation shortwards of 280 nm for a 1 bar atmosphere but does not fully protect the surface from radiation between 200 and 220 nm for eroded oxygenic atmospheres and high stellar UV levels.

For UV active stellar models, which have similar UV-B and UV-C levels to the Sun (Fig.2), the UV surface conditions for the 1 bar oxygenic atmosphere models are similar to present-day Earth. For planets with eroded atmospheres, the planetary surfaces receive UV-B and UV-C fluxes that are an order of magnitude more biologically harmful than on present-day Earth.

For UV inactive stellar models, which have lower stellar UV flux, 1 bar oxygenic atmospheres have surface UV fluxes orders of magnitude lower than Earth. The protective ozone layer also filters out most of the harmful UV-B and UV-C wavelengths. Some UV-B radiation flux reaches the surface. However these fluxes are still low enough to be tolerated by Earth life. Eroded oxygenic atmospheres (0.1 bar Earth-composition) also show a benign UV surface radiation environment for an inactive UV star model.

For an anoxic atmosphere the lack of ozone allows UV-B and UV-C radiation to penetrate the atmosphere. For the inactive star model the stellar 200-300 nm irradiation is low, keeping the UV surface environment benign. However for UV active stars, the lack of a sheltering ozone layer leads to high UV-C levels that would be lethal to most life as we know it within minutes to hours of exposure (see e.g Clark 1998). Note that certain radiation-tolerant species have demonstrated an ability to survive full solar UV in space exposure experiments (e.g. Sancho et al. 2007, Onofri et al. 2012), but they achieve this by entering a dormant state.

However, even the harsh UV surface environments for abiotic or eroded atmospheres around an active star model are comparable to model UV levels on the early Earth about 3.9 Gyr to 2.8 Gyr ago (Rugheimer et al. 2015b). Despite harsh UV-B and UV-C surface radiation prebiotic chemistry and early life flourished on the early Earth. Life could shelter from harsh UV radiation subsurface, e.g. in an ocean or under soil (see e.g. Ranjan & Sasselov 2016), but this would make it harder to detect life remotely. However, some biological UV-protection methods, such as biofluorescence, could make such a biosphere in high surface UV environments detectable (O'Malley-James & Kaltenegger 2016).

## 4 CONCLUSIONS

The UV surface habitability of planets in the HZ of the TRAPPIST-1 system depends critically on the activity of the star and the planet's atmospheric composition. The stellar UV environment will also influence the detectable atmospheric features, including any detectable biosignatures. The detection of atmospheric ozone would indicate that a planet is more likely to have habitable UV surface environments. Analysis of the originally announced

TRAPPIST-1 planets was found to be possible with JWST, which could detect present-day Earth ozone levels after 60 transits of the original innermost planet and 30 transits of original outermost planets (Barstow & Irwin 2016). Hence, future searches for ozone, combined with spectral observations of the host star's activity in the 100-400 nm range will enable us to constrain the radiation environments of the HZ planets around TRAPPIST-1, and their surface UV habitability.

### ACKNOWLEDGEMENTS
The authors acknowledge funding from the Simons Foundation (290357, LK). We thank an anonymous referee for helpful comments and suggestions.

## REFERENCES
Barstow J. K., Irwin P. G. J., 2016, MNRAS, 461, L92
Bolmont E. et al., 2017, MNRAS, 464, 3728
Bourrier V. et al., 2017, Astron Astrophys 599, L3
Buccino A. P., Lemarchand G A., Mauas P. J. D., Icarus, 192, 582
Calkins J., Barcelo J. A., 1982, The Role of Solar Ultraviolet Radiation in Marine Ecosystems. Springer US, p.143
Clark B. C., 1998, J Geophys Res, 103, 28545
De Wit J. et al., 2016, Nature, 537, 69
Demory B. O., Queloz D., Alibert Y., Gillen E., Gillon, M., 2016, ApJ Lett, 825, L25
France, K., et al., 2013, ApJ, 763, 149
Gillon M. et al., 2016, Nature, 533, 221
Gillon M. et al., 2017, Nature, 542, 456
Grant R. H., Heisler G. M., 1997), Journal of Applied Meteorology, 36, 1336
Kaltenegger L., Traub W. A., Jucks K. W., 2007, Astron J, 658, 598
Kaltenegger L. et al., 2010, Astrobiology, 10, 89
Kaltenegger L., Sasselov D, 2010, ApJ, 708, 1162
Kaltenegger L., Haghighipour N., 2013 ApJ, 777, 165
Kasting J. F., Holland H. D., Pinto J. P., 1985, J. Geophys Res Atmospheres, 90, 10497
Kasting J. F., Ackerman T. P., 1986, Science, 234, 1383
Kasting J. F., Whitmire D. P., Reynolds R. T., 1993, Icarus, 101, 108
Kopparapu R. K., Ramirez R. M., Schottelkotte J., Kasting J. F., Domagal-Goldman S., Eymet V., 2014, ApJ Lett, 787, L29
Leconte J., Forget F., Charnay B., Wordsworth R., Pottier A., 2013, Nature, 504, 268
Lindberg C., Horneck G., 1991, Journal of Photochemistry and Photobiology B: Biology 11, 699
Meadows V. S. et al., 2016, preprint (arXiv:1608.08620)
O'Malley-James J. T., Kaltenegger L., 2016, preprint (arXiv:1608.06930)
Onofri S. et al., Astrobiology, 12, 508
Parisi A. V., Downs, N., 2004, Photochemical & Photobiological Sciences, 3, 643
Ranjan S., Sasselov D. D., 2016, Astrobiology, 16, 1
Rothschild L. J., Mancinelli R. L., 2001, Nature, 409, 1092
Rugheimer S., Kaltenegger L., Segura A., Linsky J., Mohanty S., 2015a, ApJ, 809, 57
Rugheimer S., Segura A., Kaltenegger L., Sasselov D., 2015b, ApJ, 806, 137
Sancho L. G. et al., 2007, Astrobiology, 7, 443
Segura A. et al., 2005, Astrobiology, 5, 706
Segura A., Meadows V. S., Kasting J. F., Crisp D., Cohen M., 2007, Astron. Astrophys., 472, 665
Segura A., Walkowicz L. M., Meadows V. S., Kasting J. F, Hawley S., 2010, Astrobiology 10, 751
Setlow J. K., Boling M. E., 1965, Biochimica et Biophysica Acta, 108, 259
Schindler T. L., Kasting J. F., 2000, Icarus, 145, 262
Sullivan P. W. et al., 2015, ApJ, 809, 77
Toon O. B., McKay C. P., Ackerman T. P., Santhanam K., 1989, J. Geophys. Res.: Atmospheres, 94, 16287
Wheatley P. J., Louden T., Bourrier V., Ehrenreich D., Gillon M., 2017, MNRAS Lett., 465, L74